# The model of circular and longitudinal smooth muscles motions in the motility of intestinal segment

Garri Davydyan

Appletree Medical Group, 1902 Robertson Rd.,Ottawa, Ontario, K2H 7B3 Canada

**Abstract**. Coordination in circular and longitudinal muscle motions are of crucial importance in the motor function of gastrointestinal (GI) tract. Intestinal wall motions depend on myogenic-active properties of smooth muscles layers of intestinal wall, which is the ability to create active contractile forces in response to distension. Considering the stress in the circular and longitudinal smooth muscles as a sum of passive, depending on muscle deformations, and active, depending on muscle tone, components, and also assuming that the change in the muscle tone depends on the current stress-strain condition, the system of four ordinary differential equations (ODE) is obtained which describes filling-emptying cycle of intestinal segment as a process of coordinated activities of circular and longitudinal muscles of intestinal wall. A general approach in formulating the modelling conditions is based on the previously described model restricted to the circularly distensible reservoir of constant length. Obtained results illustrate the character of coordinated activities of two orthogonal muscle layers, which are alternating phases of reciprocally and uniformly changing modalities such as stretching of the wall and muscle tone. The results also contribute to the existing understanding of the roles of Auerbach and Meisner's intermuscular and submucous neural plexuses in regulations of autonomous intestinal motility, as well as clarify functional mechanisms of the interstitial cells of Cajal (ICC) in triggering of smooth muscle contractions.

**Introduction.** Under the normal physiological conditions gastrointestinal (GI) motility provides accumulation and emptying of the intestinal content. The character of the movements of chime depends on the activity of smooth muscles of the intestinal wall, which include coordinated contractions and relaxations of intestinal longitudinal and circular muscular layers. The mechanism coordinating the activities of the muscular layers of the intestinal wall is not understood. Previously, using the system of two ordinary differential equations (ODE), we described filling-emptying cycle of an intestinal segment assuming the circular muscles as the only functional component of the model. The equations describing changes in the strain-stress conditions ($\varepsilon$, $N$) of the reservoir wall have the view [1]:

$$d\varepsilon/dt = 1/2V_0 \{ P_+/z_+ + P_-/z_- - [(E\varepsilon + N)/(2\varepsilon + 1)] h_0/R_0 z_0 \} \equiv \Psi(\varepsilon, N)$$

$$dN/dt = -k_3 N^3 + k_2 N^2 - k_1 N_1 + \aleph\varepsilon + k_0(t) \equiv \Phi(\varepsilon, N), \quad 1/z = 1/z_+ + 1/z_-$$

(*)

Variables $\varepsilon$ and $N$ are circular stretching and active stress in the wall of the distensible cylinder of constant length $L_0$ and variable radius $R$ with constant pressures $P_+$ and $P_-$ and impedances $z_+$ and $z_-$ of the inlet and outlet tubes at the ends of the system. The stress-strain condition of the reservoir wall were assumed to be a sum of the passive ($E\varepsilon$), ($N$), depending on the muscle tone, components.

Although the qualitative analysis of the system helps to explain some physiologic phenomena such as the dependence of propulsive (active) contractions from the threshold values of the smooth muscle stretching and tone [5], fixed reservoir length restricts the the model to the radial wall movements and prevents to show relations with longitudinal stretching (deformations) [2, 3, 4].

The similarities of the activation and relaxation mechanisms of smooth muscle elements allow to define additional conditions for longitudinal deformations of intestinal segment, remaining, at the same time, within the frames of the main assumptions of the previous model [1].

**Statements of the model.** As a physical analogue of intestinal segment consider distensible elastic tube (reservoir) having variable radius R and length L. The reservoir wall has myogenic properties, i.e. the ability to shrink (contract) after its stress-strain condition reaches some threshold. There are two rigid tubes, inlet and outlet, attached to the both ends of the reservoir. Both tubes contain valves with impedance z±. The pressures at the ends of the system are considered to be P±. The ends of the reservoir can move along the reservoir's axis and also can change their radii remaining rigid. Conditions when input and output valves are open and closed are defined as the threshold odds in the pressures inside the reservoir and the tubes [1]. The system is a non-inertial (Fig 1).

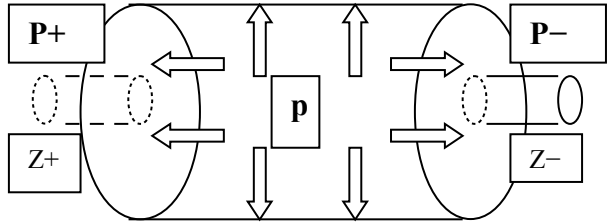

**Fig 1.** Distensible cylinder of variable length L and radius R with inlet and outlet tubes containing valves is considered as a physical model of intestinal segment. P± , z± are pressures and impedances at the ends of the system. p- pressure inside the reservoir. $(P_+ - p) > p^*_+$, $(p - P_-) > p^*_-$ are the conditions when inlet and outlet valves are open. $p^*_+$, $p^*_-$ are inlet and outlet threshold pressure drops considered constants.

In some conditions when differences in pressures make the input valve open $(P_+ - p) > p^*_+$, and output one closed $(p - P_-) \leq p^*$, the filling of the reservoir begins. Filling of the reservoir will cause the elevation in the intraluminal pressure due to reactive forces caused by circular and longitudinal distensions of the wall. It is assumed that besides the differences in pressures at the ends of the system change in the sizes of the reservoir also depends on pressure drops within the reservoir due to the differences in the radial and longitudinal forces. It becomes more demonstrative when both valves are closed. Different contractile efforts of circular and longitudinal muscles will cause the pressure drops within the reservoir and redistribution of the contents until radial and longitudinal pressure components are aligned. Independently on whether or not equilibrium is achieved in the closed reservoir, its filling or emptying will depend on the pressure drops exceeding the threshold levels at the ends of the system. Because the reservoir wall has myogenic-active properties, reaching the threshold conditions during reservoir filling will initiate active contractions of longitudinal or/and circular muscles and evacuation of the contents into the outlet or inlet tubes, depending on the conditions of the valves. After the emptying and reaching the equilibrium state a new cycle may begin [5]. If the outlet valve is always closed, the contents may undergo undulating movements until the equilibrium is reached [1].

Conditions determining the character of relationships between longitudinal and circular muscular fibers are defined in the frames of basic assumptions of the first model (*) described in [1].

Considering a reservoir wall being thin enough, stretching forces will be distributed along the longitudinal axis and circularly. For anisotropic thin shells (transversely isotropic materials), if tangential forces are considered zero, conditional equations are:

$$\sigma_1 = E_1\varepsilon_1 + (\nu_2 E_1/ E_2) \sigma_2 \qquad (1.1)$$

$$\sigma_2 = E_2\varepsilon_2 + (\nu_1 E_2/ E_1) \sigma_1 \qquad (1.2)$$

where $\sigma_1$ - circular stress in the reservoir wall, $\sigma_2$ - longitudinal stress in the wall.
$\varepsilon_1 = 1/2[(R/ R_0)^2 -1]$ - circular stretching of the reservoir wall as a mean value through the wall thickness,
$\varepsilon_2 = (L-L_0)/L_0$ – longitudinal stretching of the reservoir wall.
$E_1, E_2$ – are the Young's modules in transverse and longitudinal directions.
$\nu_1, \nu_2$ – are the Poisson's ratios.

From (1.1) and (1.2) and $E_1\nu_2 = E_2\nu_1$ we have

$$\sigma_1 = E_1(\varepsilon_1 + \nu_2\varepsilon_2)/(1 - \nu_1\nu_2) \qquad (1.3)$$

$$\sigma_2 = E_2(\varepsilon_2 + \nu_1\varepsilon_1)/(1 - \nu_1\nu_2) \qquad (1.4)$$

Consider a stress of the reservoir wall as a sum of passive $\sigma_i$ and active $N_i$ components [1]:

$$\sigma'_1 = \sigma_1 + N_1 \qquad (1.5)$$
$$\sigma'_2 = \sigma_2 + N_2 \qquad (1.6)$$

From (1.3)-(1.6) we have
$$\sigma'_1 = E_1(\varepsilon_1 + \nu_2\varepsilon_2)/(1 - \nu_1\nu_2) + N_1 \qquad (1.7)$$

$$\sigma'_2 = E_2(\varepsilon_2 + \nu_1\varepsilon_1)/(1 - \nu_1\nu_2) + N_2 \qquad (1.8)$$

Equations of reservoir's equilibrium states for circular and longitudinal forces are

$$p_1 = \sigma'_1 h/R \qquad (1.9)$$
$$p_2 = 2\sigma'_2 h/R \qquad (1.10)$$

where $p_1, p_2$ are pressures corresponding to circular and longitudinal strains of the wall, taken as mean values relatively to the wall thickness.
Considering the reservoir wall non squeezable, $RLh=R_0L_0h_0$, from (1,09), (1.10) and (1.7), (1.8) finally we obtain

$$p_1 = [E_1(\varepsilon_1 + \nu_2\varepsilon_2) + (1- \nu_1\nu_2)N_1]h_0/(1- \nu_1\nu_2)(2\varepsilon_1 + 1)(\varepsilon_2+1)R_0 \qquad (1.11)$$

$$p_2 = 2[E_2(\varepsilon_2 + \nu_1\varepsilon_1) + (1- \nu_1\nu_2)N_2]h_0/(1- \nu_1\nu_2)(2\varepsilon_1 + 1)(\varepsilon_2+1)R_0 \qquad (1.12)$$

In our case (cylindrical reservoir) longitudinal forces will be applied to the both ends of the cylinder, hence are perpendicularly to the radial forces (Fig.1).
In the static equilibrium of the reservoir we have $p_1 = p_2$. Because of the assumed possibility of redistribution of the contents, in general, $p_1 \neq p_2$.
Due to circular and longitudinal directions of the reservoir stretching, the total flow of the contents can be presented as a sum of longitudinal and radial components

$$dV/dt = dV_1/dt + dV_2/dt \qquad (2.1)$$

$dV_1/dt$ and $dV_2/dt$ are radial and longitudinal components of the volumetric velocities, respectively. According to the law of conservation of mass, change in the reservoir volume in time is proportional to the sum of pressure drops due to the differences in external forces and forces inside the reservoir. Because of radial $dV_1/dt$ and longitudinal $dV_2/dt$ components of the flow and corresponding to them pressures inside the reservoir, $p_1$ and $p_2$, it is possible to formulate two separate equations relating circular and longitudinal deformations with each component of the flow.

For the radial component we have

$$dV_1/dt = [(P_+ - p_1)/(z_+ + z_0) + (P_- - p_1)/(z_- + z_0)] - (p_1 - p_2)/z_0 \qquad (2.2)$$

$dV_1/dt = 2V_0(\varepsilon_2 + 1)d\varepsilon_1/dt$ is a velocity of changes in reservoir volume due to radial component of the flow.

Analogously is written an equation for the longitudinal component of the flow:

$$dV_2/dt = [(P_+ - p_2)/(z_+ + z_0) + (P_- - p_2)/(z_- + z_0)] - (p_2 - p_1)/z_0, \qquad (2.3)$$

$dV_2/dt = V_0(2\varepsilon_1 + 1) d\varepsilon_2/dt$ is a velocity of changes in reservoir volume due to longitudinal component of the flow.

The right parts of the equations show that the flow directed to the sides (ends) of the reservoir is a result of the radial (longitudinal) component of total flow into the reservoir (expression in square parenthesis) minus redistribution flow within the reservoir towards the sides, $(p_1 - p_2)/z_0$, (ends, $(p_2 - p_1)/z_0$).

Physical meaning of the expressions $(p_1 - p_2)/z_0$ and $(p_2 - p_1)/z_0$ becomes clear, if it is assumed that flow velocity is proportional to the pressure drops created by circular and longitudinal strains in the reservoir wall. $z_0$ is an impedance of the reservoir, for simplicity considered constant, $z_0 = $ const.

After substituting the expressions of $p_1$ and $p_2$ from (2.2), (2.3) we obtain

$$d\varepsilon_1/dt = 1/2V_0(\varepsilon_2 + 1)[ P_+/(z_+ + z_0) + P_-/(z_- + z_0) - [E_1(\varepsilon_1 + \nu_2 \varepsilon_2) + (1 - \nu_1\nu_2)N_1]h_0/(1 - \nu_1\nu_2)(2\varepsilon_1 + 1)(\varepsilon_2 + 1)R_0(1/(z_+ + z_0) + 1/(z_- + z_0) + 1/z_0) + 2[E_2(\varepsilon_2 + \nu_1\varepsilon_1) + (1 - \nu_1\nu_2)N_2]h_0/(1 - \nu_1\nu_2)(2\varepsilon_1 + 1)(\varepsilon_2 + 1)R_0 z_0 \qquad (2.4)$$

$$d\varepsilon_2/dt = 1/V_0(2\varepsilon_2 + 1)[ P_+/(z_+ + z_0) + P_-/(z_- + z_0) - 2[E_2(\varepsilon_2 + \nu_1\varepsilon_1) + (1 - \nu_1\nu_2)N_2]h_0/(1 - \nu_1\nu_2)(2\varepsilon_1 + 1)(\varepsilon_2 + 1)R_0(1/(z_+ + z_0) + 1/(z_- + z_0) + 1/z_0) + [E_1(\varepsilon_1 + \nu_2\varepsilon_2) + (1 - \nu_1\nu_2)N_1]h_0/(1 - \nu_1\nu_2)(2\varepsilon_1 + 1)(\varepsilon_2 + 1)R_0 z_0 \qquad (2.5)$$

If $N_{1,2}$ are fixed, (2.4) and (2.5) are closed system of two ODE describing stretching and contractions of the wall of the flexible cylinder along circular and longitudinal directions. For simplicity, isobars related to the changes in the conditions of the valves are not shown, so that inlet valve is always open and outlet valve closed. The wall movements reflect filling and emptying of the reservoir through the inlet tube. Under some parameters values, a phase portrait of the system is a steady focus [7, 8] (Fig.2).

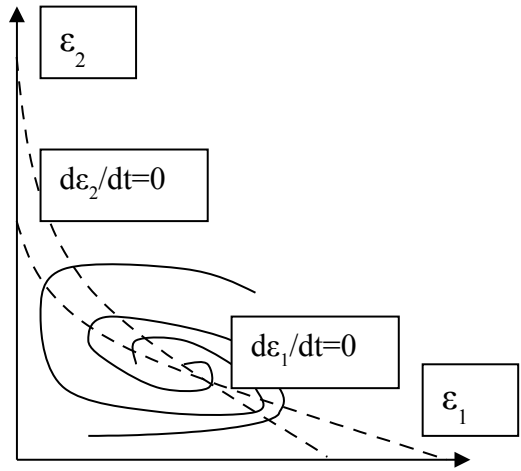

**Fig2.** **Phase curves (steady focus) on the plane ($\varepsilon_1$, $\varepsilon_2$) show alternating synphasic and reciprocal activity (stretching/contractions) of circular and longitudinal muscular layers of reservoir wall. The curves reflect alternating process of filling and emptying of the reservoir through the inlet tube. When inlet valve is open and outlet valve closed, the process fluctuates until the system reaches an equilibrium point $p_1=p_2$.**

Equilibrium state of the system (2.4), (2.5) satisfy the condition $p_1 = p_2$. There are two areas in which simultaneous contractions and stretching of circular and longitudinal muscles occur. Corresponding to the reservoir filling and emptying, these areas alternate with the ones of reciprocal activities: with the beginning of the circular contractions, first time, the longitudinal layer continues stretching until the intersection with the isocline $d\varepsilon_2/dt = 0$, after which the both layers contract simultaneously. Most probably, this evidences the fact that the increased activity of the circular muscles at first induces relaxation of longitudinal layer [2], after which contractions of the longitudinal muscles follow and vise versa. Fluctuating process with phases of relaxations and activations corresponds with the back-forward movements of the contents within the reservoir and inlet tube. Thus coordinated activities of the orthogonal layers facilitate mixing of the contents as well as propulsive forces.

In order to express velocities of active stress components $dN_1/dt$ and $dN_2/dt$ for circular and longitudinal smooth muscles, the previously stated conditions for the active stress N from [1] should be modified so that, first, to retain the ability of the smooth muscles to increase or decrease the level of its activity (tone), depending on muscle stretching $\varepsilon$, and, second, reflect the existing data, that stretching of the reservoir wall along longitudinal or circular muscle fibers, causes simultaneous contractions of reciprocal (orthogonal) layer [2, 3, 4]. The second condition means that interrelationship between $N_1$ and $N_2$ can be determined by the reciprocal character of neural regulations of circular and longitudinal muscles.

The simplest form coupling these hypothetical conditions is:

$$dN_1/dt = - k_3 N_1^3 + k_2 N_1^2 - k_1 N_1 + r(\varepsilon_1 - p\varepsilon_2) - qN_2 + k_0(t) \qquad (2.6)$$

$$dN_2/dt = - k_3' N_2^3 + k_2' N_2^2 - k_1' N_2 + r'(\varepsilon_2 - p'\varepsilon_1) - q'N_1 + k_0'(t) \qquad (2.7)$$

where $k_{0,1,2,3}$, $k'_{0,1,2,3}$, $q_{0,1,2,3}$, $q'_{0,1,2,3}$, r, r', p, p' are coefficients.

Physical meaning of ($-rp\varepsilon_{1,2} - q N_{1,2}$) is equivalent to the ionized $Ca^{2+}$ concentration in cytoplasm of the myocytes ($-q N_{1,2}$) and the flow of the $Ca^{2+}$ ($-rp\varepsilon_{1,2}$) from the surrounding tissue [10]. Minus sign tells that $Ca^{2+}$ flow to the muscle elements of one layer is accompanied by the "stealing" effect from another layer. (2.6), (2.7) are the simplest form to express the assumption of a competitive consumption of the ionized calcium of the orthogonal muscle layers.

On the plane of two variables $N_1$ and $N_2$ isoclines of the system (2.6), (2.7) are symmetric cubic parabolas with symmetry axes having negative inclination to the abscissa $N_1$. Formally this system may have from one to three stationary points. The simplest condition is the one with single equilibrium point in the area of permissible values of $N_1$ and $N_2$. The phase curves are the steady focuses (Fig 3). It is easy to see, that isoclines divide the plane on four areas, in which there is either reciprocal or unidirectional, synphasic activities, i.e. simultaneous increasing (decreasing) of the active stress components of the orthogonal layers.

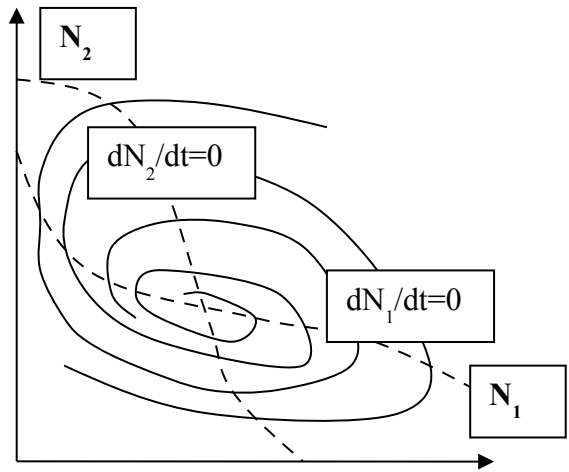

**Fig 3.** Phase curves on the plane ($N_1$, $N_2$) are steady focus with equilibrium point at the intersection of isoclines $dN_1/dt=0$, $dN_2/dt=0$, which are cubic parabolas. The character of trajectories shows alternating phases of activation/relaxation of muscle layers with reciprocal and unidirectional activities. For simplicity, $\varepsilon_1$, $\varepsilon_2$, $k_0(t)$ are constants, so that the phase trajectories reflect isovolumetric process.

The phase portraits of the system (2.4), (2.5), (2.6), (2.7) on the subspaces ($\varepsilon_1$, $N_1$), ($\varepsilon_2$, $N_2$), are qualitatively similar to those from [1]: in the conditions when inlet valve opened and outlet valve is closed (Fig.4-5), the process described are filling-emptying trajectories, where after reaching the threshold conditions (intersection with P+ isobar) spontaneous increase in the muscle tone $N_1$, $N_2$ causes evacuation of the contents into the inlet tube.

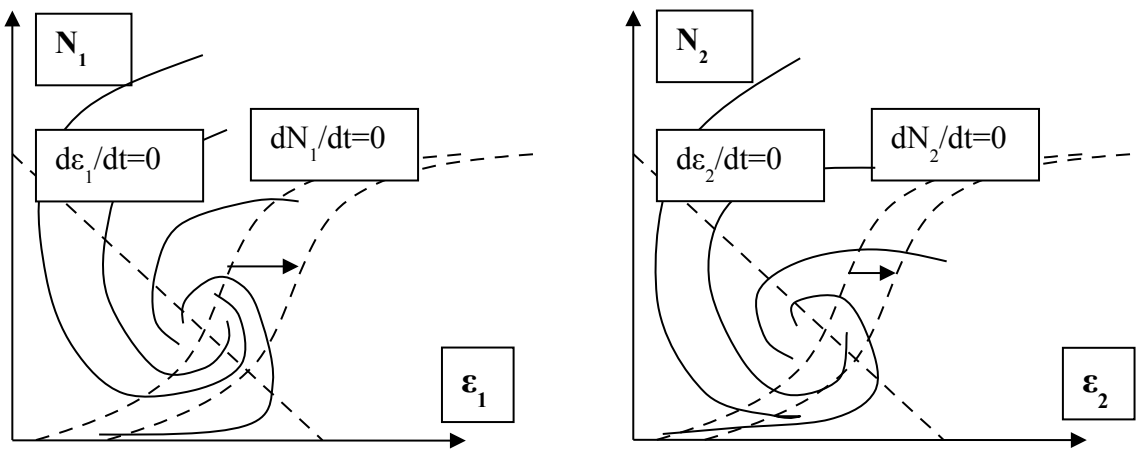

**Fig 4-5.** Phase curves of the system on the planes ($\varepsilon_1$, $N_1$) and ($\varepsilon_2$, $N_2$) show changes in the stress-strain conditions of circular and longitudinal muscle layers of the reservoir. Both phase portraits are steady focuses. For simplicity, isobars related to the changes in the conditions of the valves are not shown, so that the phase portraits illustrate the process, when inlet valve is open and outlet valve is closed [1]. ($-rp\varepsilon_2 - q N_2$) and ($-r'p'\varepsilon_1 - q' N_1$) in (2.6), (2.7) cause displacements of the isoclines $dN_{1,2}/dt=0$ along $\varepsilon_{1,2}$ axes, which displaces equilibrium points of reciprocal layers towards higher compliance.

The members ($-rp\varepsilon_{1,2} - q N_{1,2}$) from (2.6), (2.7) displace the corresponding isoclines $dN_1/dt =0$ and $dN_2/dt =0$ along the axes $\varepsilon_1$ or $\varepsilon_2$, increasing the areas of stretching of transverse or longitudinal muscular layers (Fig 4-5).

**Conclusion.** The system of four first order ODE is a continuation of the idea to describe basic physiologic mechanisms and some functional phenomena of intestinal motility [1, 5, 6]. It gives an illustrative picture of coordinated activities of circular and longitudinal smooth muscle layers of intestinal wall. In the frame of the quite general assumptions, obtained results are agreed with the existing data and clarify mechanisms of coordinated activity of "orthogonal" muscular layers of intestinal wall [2, 3, 4]. Thus, not only does an initial level of muscle tone of one layer affect threshold stretching of the wall, but so does the stress-strain condition of its orthogonal counterpart. Alternating changes in threshold conditions of circular and longitudinal muscle layers find its physiologic benefit in undulating movements of intestinal contents, facilitating the contact with the mucous membrane and absorption of nutrients. Defining functional conditions of the intestinal wall through the muscle stretching ($\varepsilon$) and a tone ($N$), and also considering the myogenic active properties of smooth muscles (the ability to contract despite stretching forces) as depending on a tone component ($N$) in stressed muscles, are the clues in explaining the physiologic mechanism of the Starling's law.

The presented results may explain the role of the Auerbach and Meisner's intermuscular and submucous neural plexuses in autonomous regulatory mechanisms of coordinated activities of circular and longitudinal muscular layers, as well as clarify the function of interstitial cells of Cajal (ICC) in triggering of smooth muscle contractions [9].

Further analytical and physiological investigations of the model can also be helpful in determinine the influence of transverse and longitudinal sizes of surgically created reservoirs on filling-emptying function [5].

**Acknowledgments.** I would like to thank S.A Regirer for discussions and valuable comments.